# NMR Evidence for the Topologically Nontrivial Nature in a Family of Half-Heusler Compounds

Xiaoming Zhang[1,+], Zhipeng Hou[1,+], Yue Wang[1,+], Guizhou Xu[1], Chenglong Shi[1], EnKe Liu[1], Xuekui Xi[1,*], Wenhong Wang[1,*], Guangheng Wu[1], and Xi-xiang Zhang[2]

[1]State Key Laboratory for Magnetism, Beijing National Laboratory for Condensed Matter Physics, Institute of Physics, Chinese Academy of Sciences, Beijing 100190, China.

[2]Physical Science and Engineering, King Abdullah University of Science and Technology, Thuwal 23955-6900, Saudi Arabia.

[*]Correspondence and requests for materials should be addressed to X.K.X (email: xi@iphy.ac.cn) or W.H.W (email: wenhong.wang@iphy.ac.cn).

[+] These authors contributed equally to this work.

**Spin-orbit coupling (SOC) is expected to partly determine the topologically nontrivial electronic structure of heavy half-Heusler ternary compounds. However, to date, attempts to experimentally observe either the strength of SOC or how it modifies the bulk band structure have been unsuccessful. By using bulk-sensitive nuclear magnetic resonance (NMR) spectroscopy combined with first-principles calculations, we reveal that $^{209}$Bi NMR isotropic shifts scale with relativity in terms of the strength of SOC and average atomic numbers, indicating strong relativistic effects on NMR parameters. According to first-principles calculations, we further claim that nuclear magnetic shieldings from relativistic $p_{1/2}$ states and paramagnetic contributions from low-lying unoccupied $p_{3/2}$ states are both sensitive to the details of band structures tuned by relativity, which explains why the hidden relativistic effects on band structure can be revealed by $^{209}$Bi NMR isotropic shifts in topologically nontrivial half-Heusler compounds. Used in complement to surface-sensitive methods, such as angle resolved photon electron spectroscopy and scanning tunneling spectroscopy, NMR can provide valuable information on bulk electronic states.**

Since a number of heavy half-Heusler ternary compounds were proposed as a new class of three-dimensional topological insulators with a tunable nontrivial band structure as a result of strong spin-orbit coupling (SOC) [1-4], this kind of compounds have attracted considerable attentions as multifunctional quantum materials [5-7]. Their nontrivial bulk band structure with relativistic-like character has proven to be closely linked to their strong SOC [8-10], as is inferred from their unique transport properties such as extremely high carrier mobility, large linear magneto-resistance, and pronounced weak anti-localization effect [11-17]. Similar to the well-known topologically nontrivial HgTe compound [18,19], the inversion of an $s$-like $\Gamma_6$ band in heavy atom containing half-Heusler compounds, identified by numerical computations, is interesting and claimed to result from relativistic effects [20]. Meanwhile, the SOC effect, which is purely relativistic in origin[10,21], is expected to split the degenerate $p_{xyz}$-$\Gamma_5$ states (primarily being $t_{2g}$ and $p$ orbitals) into quartet $p_{xy}$-$\Gamma_8$ ($j$=3/2, denoted as $p_{3/2}$) close to the Fermi level and doublet $p_z$-$\Gamma_7$ ($j$=1/2, denoted as $p_{1/2}$) states below the Fermi level, as illustrated in Figure 1(a-c). The experimental determination of their topologically nontrivial nature is of fundamental importance for setting the foundation of their potential applications. However, experimental verification remains limited to comparing band calculations with conventional angle-resolved photoemission spectroscopy (ARPES) [7,22]. Unfortunately, likely technical challenges using this type of surface-sensitive method have precluded success in revealing the topologically nontrivial band structure nature of unstrained bulk half-Heusler compounds[22].

Because relativistic effects influence nuclear magnetic shielding, relativistic modifications of the bulk band structure should, therefore, be detectable by locally susceptible sensitive methods such as nuclear magnetic resonance (NMR) [23,24]. For example, the probability density at the nucleus of relativistic $p_{1/2}$ states is not zero [25], therefore, we can expect additional contributions to the nuclear magnetic shieldings in these bulk half-Heusler compounds [see Figure 1(d-f)]. Magnetic shieldings will not cancel out due to the different occupancy of $p_{1/2}$ tuned by relativity among the members in this family of compounds [26]. Moreover, nuclear magnetic shielding is sensitive to the change of symmetry of the conduction bands owing to the inversion of the $s$-like $\Gamma_6$ band. Although, relativistic effects including SOC on NMR chemical shifts and nuclear spin-spin interactions on small molecules and complexes [21,27,28] and on the surface states of $Bi_2Te_3$ [24] have already been widely acknowledged and discussed, it is highly desirable for systematic investigation of the relativistic effects on the NMR spectra in heavy half-Heusler intermetallics[29-31]. In addition, the underlying mechanism of the correlation between NMR parameters and topologically nontrivial nature remains inconclusive.

In this work, we demonstrate the relativistic effects, including SOC on bulk band structures in topological half-Heusler R-M-Bi (R= Sc, Y, Lu, M= Ni, Pd, Pt) compounds, by combining bulk $^{209}$Bi NMR spectroscopy with a state-of-the-art first-principles calculation. Here, $^{209}$Bi (spin $I$ = 9/2, 100% natural abundance) NMR is selected as an atomic probe because Bi is a common heavy component in these RMBi half-Heusler compounds. In addition, $^{209}$Bi NMR is one of the most sensitive atomic probes to changes in chemical shieldings at Bi sites in a cubic structure where quadruple interactions can be neglected. We find that $^{209}$Bi NMR shift tensors rescale with the relativity represented by average nuclear charge <Z>, as well as the band topology characterized by $E_{BIS}$ (defined as energy difference between $\Gamma_6$ and $\Gamma_8$ bands, $E_{BIS} = E_{\Gamma 6} - E_{\Gamma 8}$) or $E_{SOC}$ ($E_{SOC} = E_{\Gamma 8} - E_{\Gamma 7}$) in these half-Heusler compounds. Evidence presented here for relativistic effects on bulk band structures probed by $^{209}$Bi NMR provides an important insight into the mechanism of band inversion and its role on transport properties in these topologically nontrivial half-Heusler compounds.

**Results**

**Crystal growth and characterization.** Single R-M-Bi crystals were grown out of Bi flux and characterized by X-ray diffraction (XRD) with Cu K$_\alpha$ radiation on a Rigaku x-ray diffractometer. Details on growth procedures and XRD characterizations can be found in the supplemental materials. These nonmagnetic compounds crystallize in a cubic MgAgAs structure with a space group number of 216 (*F-43m*), which can be viewed as three face-centered cubic sublattices positioned at (0,0,0), (1/4,1/4,1/4) and (3/4,3/4,3/4) along the body diagonal. $^{209}$Bi NMR spectra of powdered single crystals were obtained by integrating spin-echo intensity with a Bruker Avance III 400 spectrometer in a magnetic field of ~9.39 Tesla. A Hahn-echo pulse sequence was used to record echo intensity with the last delay time around 300 ms because of the short spin-lattice relaxation time (~ 60 ms). As determined from nutation experiments, the measured typical 90 degree time ($t_{\pi/2}$) is about 5 μs for the samples, close to that of aqueous Bi(NO$_3$)$_3$ at the same *r.f.* condition. $^{209}$Bi chemical shifts were referenced to a 1M Bi(NO$_3$)$_3$·5H$_2$O aqueous nitrate solution (8.4 ppm at 298 K).

**Analysis of composition dependence of NMR spectra.** Figure 2(a) presents typical static $^{209}$Bi (spin $I = 9/2$) NMR spectra measured at room temperature on powdered single crystals of isoelectronic LuPd$_{1-x}$Pt$_x$Bi (x=0, 0.2, 0.9 and 1). Only $^{209}$Bi central lines can be observed, as expected for the cubic symmetry. The central linewidth defined as full line width at half magnitude (FWHM) is ~20 kHz for LuPtBi, which is much larger than that estimated for pure dipolar broadening (~ 1 kHz), indicating the existence of other contributions to the broadening such as second-order quadrupole, chemical-shift anisotropy or site distributions. Potentially, the atomic disorder will break the local cubic symmetry at Bi sites and lead to site distributions. Since the measured quadrupole coupling constants are very small, it is reasonable to neglect the contributions of $^{209}$Bi central line broadening and anisotropic shifts from second-order quadrupole interactions. The observed nearly symmetrical central line shape for LuPdBi within which the peak locates at the center of gravity, we can thus define $^{209}$Bi isotropic shifts as the peak position of the central lines. Meanwhile, the central line shape of LuPtBi is asymmetric and, therefore, the positions of the isotropic resonances are simulated by shift anisotropy. In addition, site distributions make no contributions to isotropic shifts.

Similar to the well-studied Bi$_{1-x}$Sb$_x$ topological insulators, the SOC strength in LuPd$_{1-x}$Pt$_x$Bi can be tuned within a certain range of $x$ [32]. Figure 2(b) reveals that $E_{SOC}$ (= $E_{\Gamma 8}$ - $E_{\Gamma 7}$) derived from band structure calculations, as an indicator of SOC strength, increases in magnitude with Pt alloying. Details of the c alculations and analysis are described in the supplemental materials. Simultaneously, $^{209}$Bi NMR resonant frequency moved toward upfield upon Pt alloying. The measured $^{209}$Bi isotropic shift is about -6300 ppm for LuPtBi and -4300 ppm for LuPdBi. Similar values for the two compounds measured at the same temperature but at a different magnetic field have been reported [31], indicating that these values are independent of magnetic field. Upon Pt alloying in LuPd$_{1-x}$Pt$_x$Bi, the $^{209}$Bi isotropic shifts change from -4830 ppm for LuPd$_{0.8}$Pt$_{0.2}$Bi, to -6059 ppm for LuPd$_{0.1}$Pt$_{0.9}$Bi. Note that Figure 2(b) shows that the calculated $E_{SOC}$ and measured $^{209}$Bi shifts are strongly correlated with each other, where Pt concentration is an implicit parameter. This result indicates that the SOC parameter is coupled to $^{209}$Bi NMR isotropic shifts in LuPd$_{1-x}$Pt$_x$Bi series.

**Scaling of $^{209}$Bi NMR isotropic shifts.** Since the relativistic modifications of band structures have a similar mechanism in these R-M-Bi compounds [1,2,10], it would be interesting to link $^{209}$Bi NMR shifts with $E_{BIS}$ or $E_{SOC}$ if possible. For a convenient comparison, Table I lists the measured $^{209}$Bi isotropic shifts, the calculated $E_{BIS}$、 $E_{SOC}$ values and their topological status. Figure 3(a) shows that rough correlations can be built between $^{209}$Bi isotropic shifts and average SOC strength in terms of $E_{SOC}$ if the R-M-Bi compounds are treated as two subgroups for consideration, where five members (ScNiBi、YNiBi、LuNiBi、LuPdBi、LuPtBi) form one subgroup, and four members (ScPdBi、YPdBi、ScPtBi、YPtBi) form the other. The average SOC strength can be also represented by average atomic number <Z> or nuclear charge. In general, the SOC constant [21], $a = \dfrac{\mu_0 Z e^2 \hbar^2}{8\pi m_e^2 r^3}$ is proportional to Z (nuclear charge number), where $r$ refers to radial distance and other symbols have their normal meanings. Therefore, nuclear magnetic shieldings will scale with Z when relativistic effects are important. Figure 3(b) clearly shows that $^{209}$Bi isotropic shifts almost linearly follow <Z> for the two subgroups, where <Z> = ($Z_R+Z_M+Z_{Bi}$)/3 in R-M-Bi compounds. With <Z> increasing from ~40 to ~80, $^{209}$Bi shifts from about +3000 ppm to -6500 ppm. These results indicate that relativistic effects are relevant for understanding the mechanism of $^{209}$Bi NMR isotropic shifts in these half-Heusler compounds. The average SOC strength is believed to contribute to the value of $E_{BIS}$ [1,20]. In Figure 3(c) we observe clear correlations between $^{209}$Bi isotropic shifts and $E_{BIS}$ in R-M-Bi compounds, especially the changing along subgroups labeled by a particular binary MBi = (PtBi, PdBi and NiBi).

To further reveal the relativistic effects on NMR shifts, a relativistic density functional approach to calculate spin-orbit shifts involving heavy $^{209}$Bi nuclei was performed at various relativistic levels. All calculations are implemented in WIEN2k code using the augment plane-wave DFT-NMR method [33-35], and calculated band structures were similar to previous reports[4]. Previously, this method was used successfully to calculate NMR chemical shifts in small molecular systems or diamagnetic semiconductors [35]. To calculate $^{209}$Bi isotropic shifts in solids containing heavy atoms, relativistic SOC effects are invoked to account for the relativistic $p_{1/2}$ electrons. Figure 3(d) shows the total calculated $^{209}$Bi chemical shieldings with SOC included. Fittings of the calculated chemical shieldings with respect to the experimental shifts yield the calculated $^{209}$Bi isotropic shifts, as practiced in [ref.35]. As shown in Figure S4, One can see that the calculated $^{209}$Bi shifts with SOC follow the measured ones more closely than the calculated ones with only scalar relativistic corrections. In other words, computations indicate that SOC plays a significant role in $^{209}$Bi NMR isotropic shifts in these heavy half-Heusler compounds.

In addition to isotropic shifts, traces of hidden relativistic effects on NMR can also be revealed through bonding properties and nuclear spin-spin interactions (up to several kHz) that are usually less than quadruple or dipole-dipole couplings in solids. As shown in Figure 2(a), the central line shape of the $^{209}$Bi powder spectra of LuPtBi compound is asymmetric. Based on symmetry analysis, dipolar interactions do not account for the shift anisotropy in cubic crystals. This anisotropy is potentially an indicator of strong nuclear spin-spin interactions originating from the phase factor for relativistic $p_{1/2}$ orbitals [36]. The relativistic increase of the shift anisotropy seems to be a general phenomenon that has also been found in small molecular systems [21]. Nuclear spin-spin coupling constants between 6$^{th}$-row elements have been predicted to increase by almost one order of magnitude in comparison with their 5$^{th}$-row counterparts, in line with the well-known relativistic bond contraction. For instance, nuclear spin-spin coupling constant can reach as high as 57 kHz for Pt-Tl [21]. Meanwhile, the central line width of a $^{209}$Bi NMR spectrum taken at room temperature is ~20 kHz. This line width is free of 2$^{nd}$-order quadrupolar broadening and almost one order of magnitude larger than that from nuclear spin dipolar interactions (~ 1 kHz), as discussed earlier, indicative of the line broadening contribution from strong nuclear spin-spin coupling. Enhanced nuclear spin-spin coupling is also in line with the well-known relativistic bond contraction, which is perfectly consistent with our experimental observations. LuPtBi is found to be stoichiometric and free of antisites, as indicated by XRD refinement. Within the lattice, M-Bi forms a covalent-type ZnS substructure [7] from orbital hybridization, and the lack of antisites suggests that the Lu-Pt-Bi bonds are stronger than their counterparts Lu-Pd-Bi.

**Discussion**

As the strength of band inversion which characterizes band topology has already been found to be connected with <Z> [1,2], the scaling of the magnitude of NMR isotropic shift tensors and spin-lattice relaxation times with relativity provides new perspective into the bulk band topology of these half-Heulser compounds. The $^{209}$Bi isotropic shift tensor seems to follow a simple rule [29-31]: the theoretically predicted nontrivial phases display nearly zero or negative $^{209}$Bi shifts, while trivial phases show large positive shifts (see Table I). For example, the measured $^{209}$Bi shifts at room temperature are -6362 ppm for LuPtBi, -4368 ppm for LuPdBi and -1748 ppm for YPtBi whereas 1200 ppm for YPdBi, 2608 ppm for ScNiBi and 3188 ppm for ScPdBi. To be noted, $^{209}$Bi NMR shifts extrapolated to zero temperatures exhibited only minor changes in magnitude compared to those taken at room temperature (except for LuPtBi for unkown reasons, see Figure S3). In addition, there is no change in sign when extrapolated to zero temperatures for all the compounds. From the scaling of the magnitude of $^{209}$Bi shifts with relativity, it can now be understood that the change of sign in NMR shifts is due to the competitions among spin orbit shifts ($K_{SO}$) induced by relativistic SOC, scalar-relativistic shifts ($K_{Scalar}$) and non-relativistic shifts ($K_{NR}$) from respective $s$ or $p/d$ electrons within the vicinity of the Fermi level. Total NMR shifts can be described as: $K_{total} = K_{NR} + K_{Scalar} + K_{SO}$ [21]. Since the hyperfine field of $s$ character electrons (MHz) is usually orders of magnitude of larger than that of $p/d$ electrons (kHz), even the trace of $s$ states at the Fermi level will contribute to the total hyperfine magnetic shift via the direct Fermi-contact mechanism, which is positive just like in metals. When the $s$-like $\Gamma_6$ band is inverted and far away from the Fermi level, however, the positive Fermi-contact contribution will be lost. Furthermore, as mentioned earlier, in contrast to $p_{3/2}$ states, the probability density of relativistic $p_{1/2}$ electrons, $\left\langle \left| \Psi_{p_{1/2}} \right|^2 \right\rangle$ does not vanish under strong spin-orbit modification in heavy atoms [25]. A net spin density in the ground state will then be induced by an external field because of SOC. Assuming the SOC-induced net spin density from $p_{1/2}$ electrons has a spin component opposite to the external magnetic field for a topologically nontrivial half-Heusler compound, an internal magnetic field opposite to the external field at the Bi nucleus will be produced. $^{209}$Bi NMR shifts will become increasingly negative, as the occupancy of $p_{1/2}$ states becomes larger, as shown in computations (see Figure 3d and Figure 4).

Based on our NMR data we draw the conclusion that the occupancy of $p_{1/2}$ states varies depending on their position controlled by SOC, which will affect the measured NMR shifts. Details of the calculations are described in the Methods section.

In Figure 4, we show the electronic bulk band structures under fully relativistic calculations for (a) ScNiBi, (b) ScPdBi, (c) LuPdBi and (d) LuPtBi, respectively, where the degree of occupancy of $p_{1/2}$-like ($\Gamma_7$) bands from Bi contributions are denoted as different sizes of red dots. The calculated band structures are in good accordance with former results[4]. During the period from (a) ScNiBi to (d) LuPtBi, we can infer that firstly, the SOC effect in the system is enhancing and pushes $p_{1/2}$ states lower, away from the Fermi level and that secondly, the occupancy of relativistic $p_{1/2}$ states from Bi contributions increases with SOC strength. Such differences in orbital occupancy and energy position originating from the SOC effect will affect the magnitude of $^{209}$Bi nuclear shieldings via a Fermi-contact-like mechanism [26]. Therefore, SOC will affect the NMR shift by tuning the relative occupancy of $p_{1/2}$ states, which is at the heart of relativistic $p_{1/2}$ shielding mechanism.

In conclusion, we have found the correlation of $^{209}$Bi shift tensors with relativistic SOC strength in terms of $E_{SOC}$ and $<Z>$ in a class of half-Heusler R-M-Bi compounds by using NMR spectroscopy. The topologically nontrivial nature in the half-Heusler compounds is manifested as large differences in the magnitude of $^{209}$Bi NMR isotropic shifts between trivial and nontrivial half-Heusler members. This finding provides new insight into the mechanism of relativistic effects on band inversion and its role on transport properties in the topologically nontrivial half-Heusler compounds. Additional technical improvements in the NMR measurements may be directed towards understanding how the relativistic SOC effects on electronic structure manifest themselves in NMR shifts.

**Methods**

**Sample growth.** Single crystals of half-Heusler R-M-Bi (R= Sc, Y, Lu, M= Ni, Pd, Pt) were grown by using the self-flux method. The high-purity starting materials Sc, Y, Lu (ingot, 99.99%), Ni, Pd, Pt (ingot, 99.99%) and Bi (ingot, 99.99%) were mixed together in a molar ratio of 1:1:10, and afterward the mixture of pure elements were placed in an alumina crucible. This procedure was performed in a glove box filled with Ar gas, where the oxygen and humidity content was less than 0.5 ppm. The mixture was sealed inside a tantalum tube. This procedure was operated in a home-made arc furnace. The sealed tantalum tube was then sealed into an evacuated quartz tube. Crystal growth was performed in a furnace by heating the tube from room temperature up to 1150 over a period of 15 h, and maintained for 24 h before slowly cooling down to 650 at a rate of 2 /h. The excess Bi flux was removed at 650 by spinning the tube in a centrifuge at rate of ~3000 r/min. After the centrifugation process, most of the excess flux was removed from crystal surfaces and the remaining flux was etched in a diluted hydrochloric acid (See Supplementary materials).

**Structural characterization.** The composition of the single crystal samples was determined by energy-dispersive X-ray (EDX) spectroscopy and the structure were checked by X-ray diffraction (XRD) with Cu-$K\alpha$ radiation. The single-crystal orientation was checked by a standard Laue diffraction technique (See Supplementary materials).

**NMR measurements and data analysis.** $^{209}$Bi NMR spectra of powdered single crystals were obtained by integrating the spin-echo intensity with a Bruker Avance III 400 HD spectrometer in a magnetic field of ~9.39 Tesla. Hahn echo pulse sequence was used for recording echo intensity. The first pulse length (~5 μs) is about 1/2 of the 90 degree time ($t_{\pi/2}$) of the samples to prevent distortion of Fourier transformed spectra. $^{209}$Bi chemical shifts were referenced to Bi(NO$_3$)$_2$ aqueous nitrate solution (Bi(NO$_3$)$_3$5H$_2$O in 1 M nitrate solution, 8.4 ppm) (See Supplementary materials).

**Band structure calculations.** All calculations are implemented in WIEN2k code using the augment plane-wave DFT-NMR method and the band structure calculations were performed using the full-potential linear-augmented plane wave code implemented in the WIEN2K package [33, 34]. The converged ground state was obtained using 10 000 k points in the first Brillouin zone. A combination of a modified Becke-Johnson exchange potential and the correlation potential of the local-density approximation were used to obtain the band structures [37], as it predicts band gaps and the band order with favorable accuracy [4]. Spin-orbit coupling (SOC) was treated as a second variational procedure with scalar-relativistic orbitals as a basis, where states up to 10 Ry above the Fermi level were included in the basis expansion. Chemical shielding and EFG tensors were calculated using density functional theory (DFT), as implemented in NMR- WIEN2K [34, 35], where both cases for scalar and fully relativistic effective core potentials were considered.

**Acknowledgements**

This work was supported by the National Program on Key Basic Research Project (Grant No. 2012CB619405 and 2011CB012800), National Natural Science Foundation of China, NSFC (Grant Nos. 51371190, 51301195 and No. 11474343), and Strategic Priority Research Program B of the Chinese Academy of Sciences under the grant No. XDB07010300.

**Author contributions**

W.H.W. and X.K.X. designed the research project. Z.P.H, Y.W., G.Z.X., C.L.S. and X.M.Z. performed all the experimental measurements. X.M.Z. and E.K.L. performed band structure and NMR calculations. G.H.W. and X.X.Z. helped with the results analysis. X.M.Z., X.K.X. and W.H.W. wrote the manuscript. All authors reviewed the manuscript.

**Additional information**

Supplementary information accompanies this paper at http://www.nature.com/

Scientificreports

Competing financial interests: The authors declare no competing financial interests.

Figures and Figure captions:

**Figure 1 | Crystal structure and bulk electronic states of Half-Heusler R-M-Bi.** Schematic diagrams of (**a**) the crystal structure, (**b**) bulk electronic band states without SOC (denoted as W/S), and (**c**) with SOC, respectively. For the band structures, SOC splits the $p_{xyz}$-$\Gamma_5$ state into $p_{xy}$-$\Gamma_8$ (j=3/2, denoted as $p_{3/2}$) and $p_z$-$\Gamma_7$ (j=1/2, denoted as $p_{1/2}$) states. The SOC parameter is defined as $E_{SOC}$ = $E_{\Gamma 8}$ - $E_{\Gamma 7}$, where a larger $E_{SOC}$ shows a stronger SOC strength in system. An illustration of SOC effects on NMR chemical shifts (**d**), (**e**) and (**f**). The SOC-split $p_{1/2}$ state imposes an additional magnetic shielding to the nucleus through hyperfine interactions. The direction of shielding tensor from SOC effects can be different, depending on the nature of exchange interactions.

**Figure 2 | $^{209}$Bi NMR spectra in isoelectronic alloys LuPd$_{1-x}$Pt$_x$Bi (x=0, 0.2, 0.9, 1).** (**a**) Typical $^{209}$Bi powder spectra of half-Heusler alloys LuPdBi, LuPd$_{0.8}$Pt$_{0.2}$Bi, LuPd$_{0.1}$Pt$_{0.9}$Bi and LuPtBi. All spectra are normalized with respect to peak intensity. The peak positions are referenced to Bi(NO$_3$)$_2$ in aqueous solution. (**b**) The measured NMR $^{209}$Bi isotropic shifts and the calculated SOC parameter $E_{SOC}$ for the half-Heusler system LuPd$_{1-x}$Pt$_x$Bi.

**Figure 3 | Scaling of $^{209}$Bi isotropic shifts.** The measured NMR $^{209}$Bi isotropic shifts as a function of (**a**) the calculated SOC parameter $E_{SOC}$, (**b**) the average nuclear charge <Z> (=($Z_R$+ $Z_M$+ $Z_{Bi}$) /3), (**c**) the calculated band inversion strength $E_{BIS}$ and (**d**) the calculated NMR $^{209}$Bi isotropic shifts.

**Figure 4 | Electronic bulk band structures under fully relativistic calculations for (a) ScNiBi, (b) ScPdBi, (c) LuPdBi and (d) LuPtBi.** The size of the red dots indicates the degree of occupancy of $p_{1/2}$-like ($\Gamma_7$) bands from Bi contributions. The evolution of orbital occupancy and energy position with SOC strength will affect the magnitude of $^{209}$Bi nuclear shieldings via a Fermi-contact-like mechanism.

**Table I** The measured values of lattice constants, NMR $^{209}$Bi isotropic shifts at room temperature, calculated band inversion strength $E_{BIS}$ (=$E_{\Gamma 6}$-$E_{\Gamma 8}$), SOC parameter $E_{SOC}$ (=$E_{\Gamma 8}$-$E_{\Gamma 7}$) and the topological status for half-Heusler R-M-Bi alloys. The values of lattice constants are accordance with the literature.[1-4,12-15]

| Alloys | Lattice constant (Å) | Isotropic shift (ppm 300K) | Calculated $E_{BIS}$ (eV) | Calculated $E_{SOC}$ (eV) | Topological status |
|---|---|---|---|---|---|
| ScNiBi | 6.19 | 2608 | 1.651 | 0.218 | No |
| ScPdBi | 6.43 | 3188 | 0.721 | 0.369 | No |
| ScPtBi | 6.50 | 382 | -0.862 | 0.616 | Yes |
| YNiBi | 6.41 | 900 | 1.130 | 0.334 | No |
| YPdBi | 6.63 | 1200 | 0.303 | 0.501 | No |
| YPtBi | 6.64 | -1748 | -0.864 | 0.686 | Yes |
| LuNiBi | 6.34 | -2223 | 0.506 | 0.383 | No |
| LuPdBi | 6.57 | -4368 | -0.496 | 0.498 | Yes |
| LuPtBi | 6.58 | -6362 | -1.337 | 0.701 | Yes |

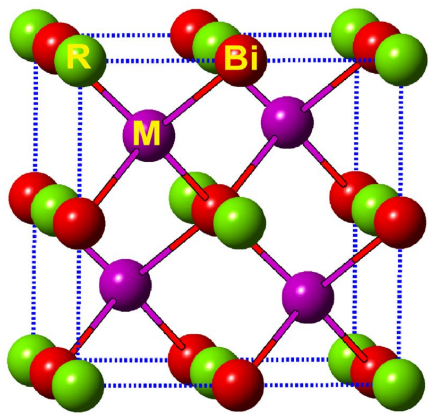
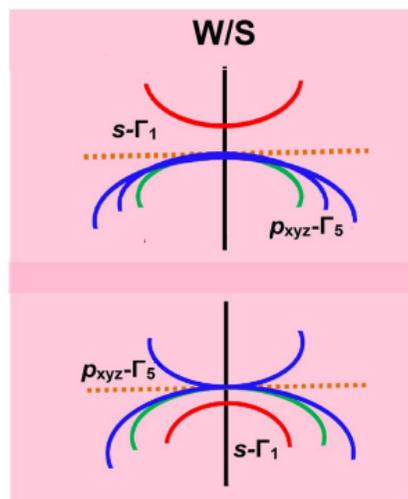
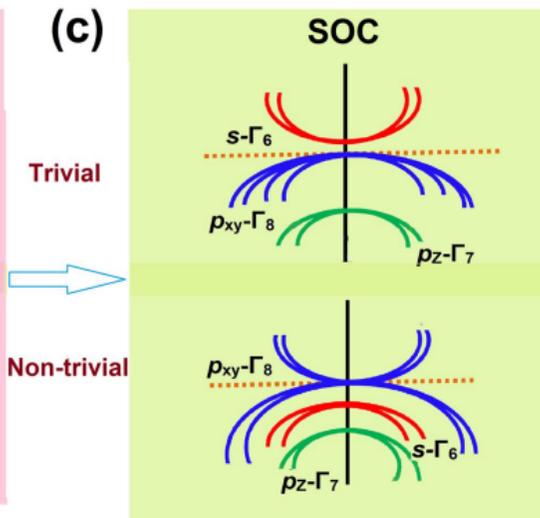
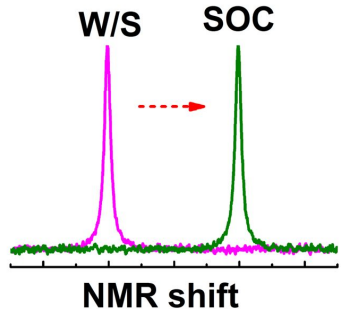
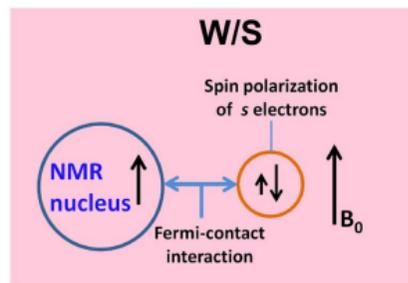
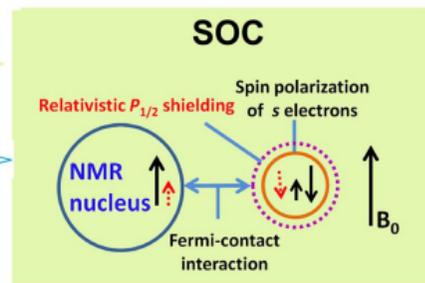

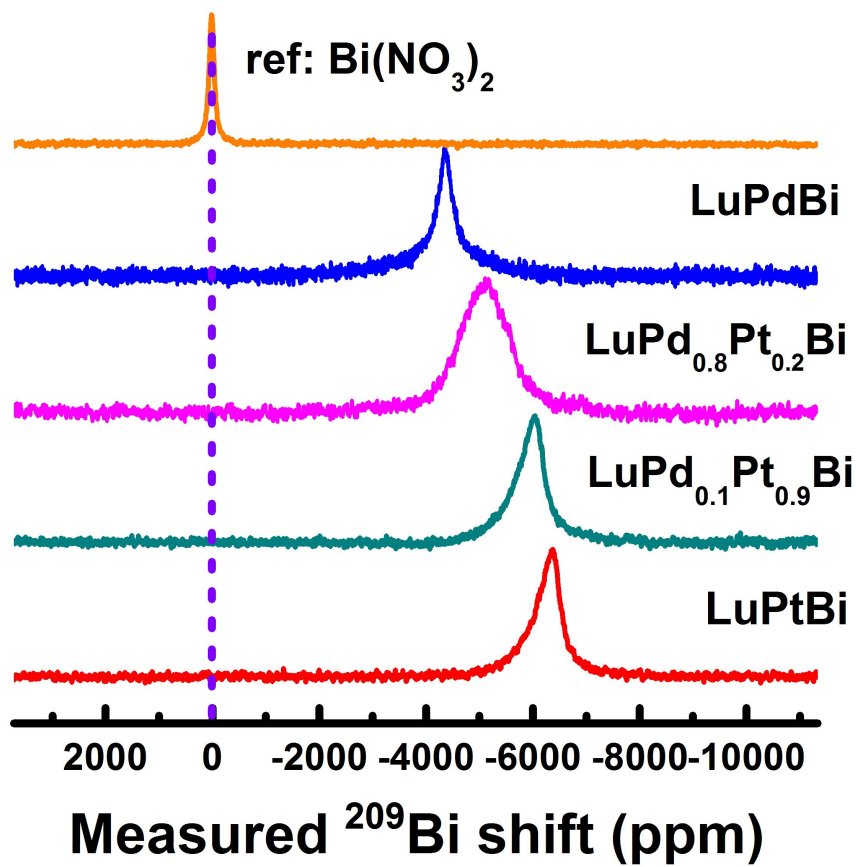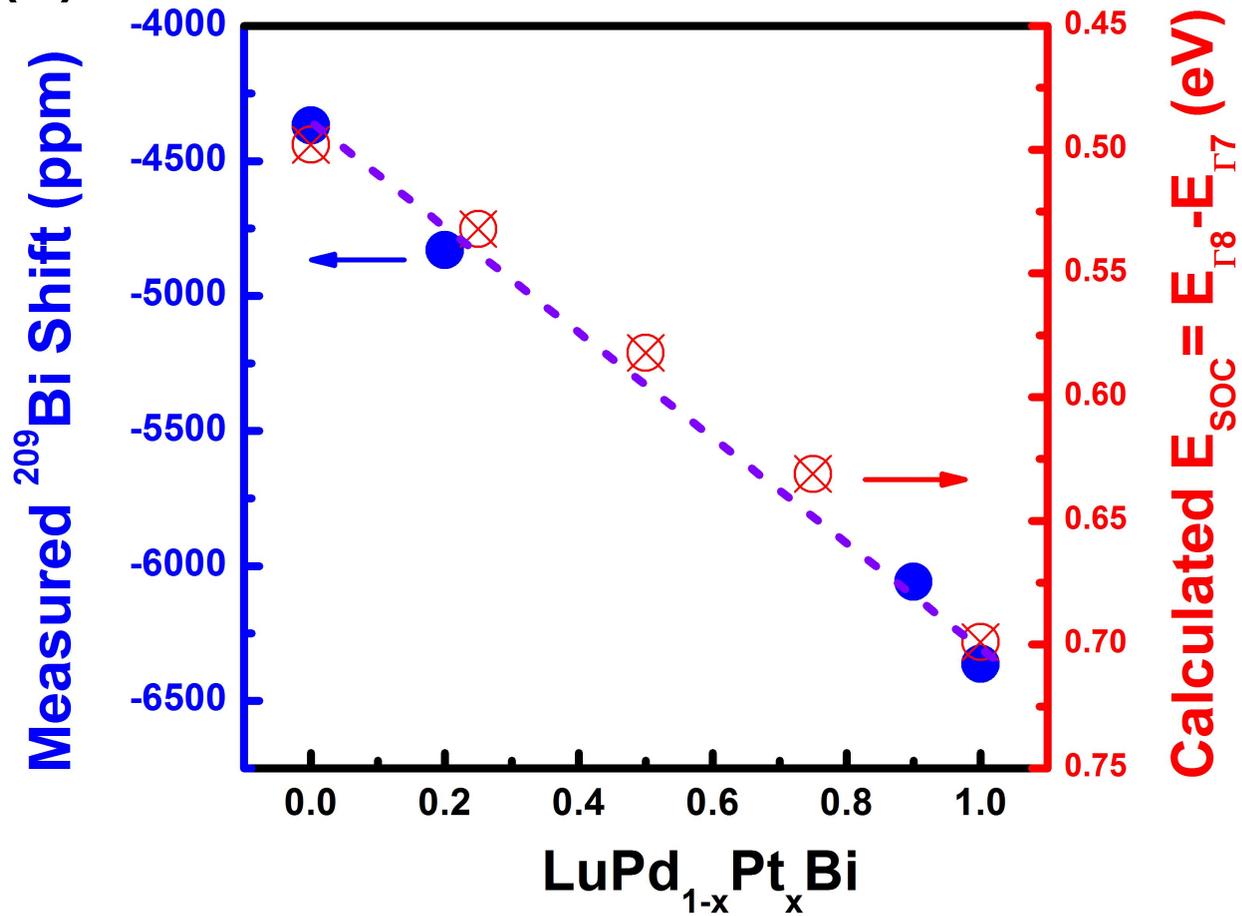

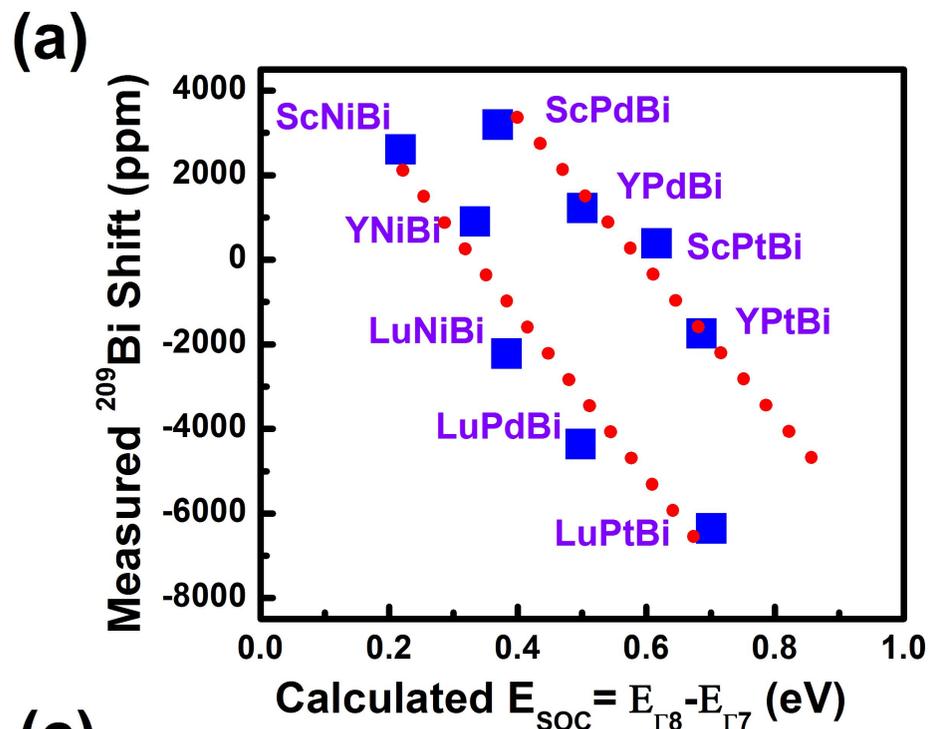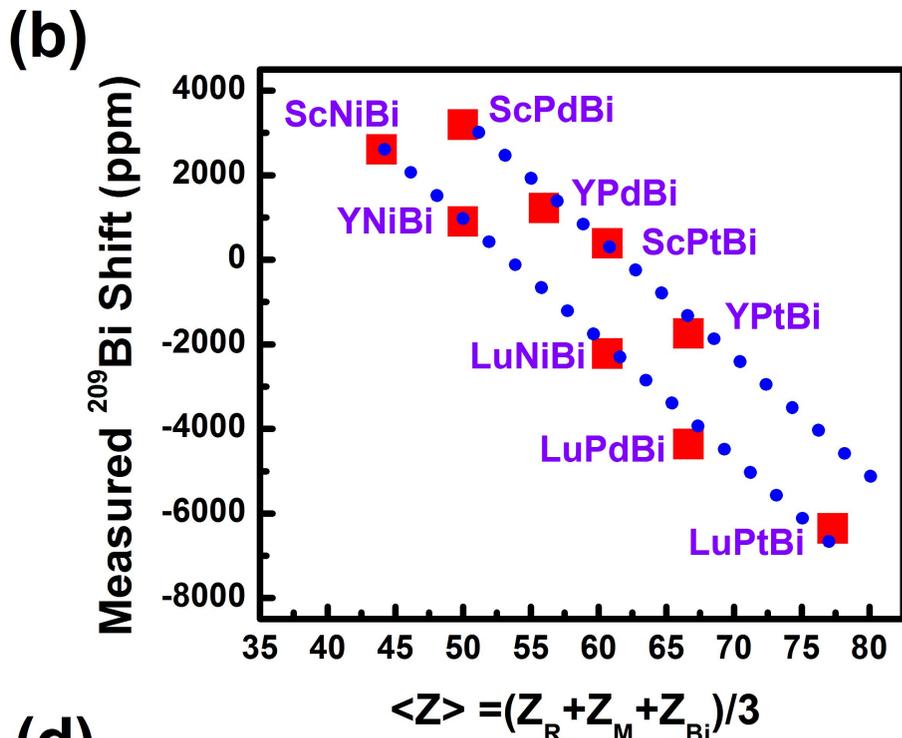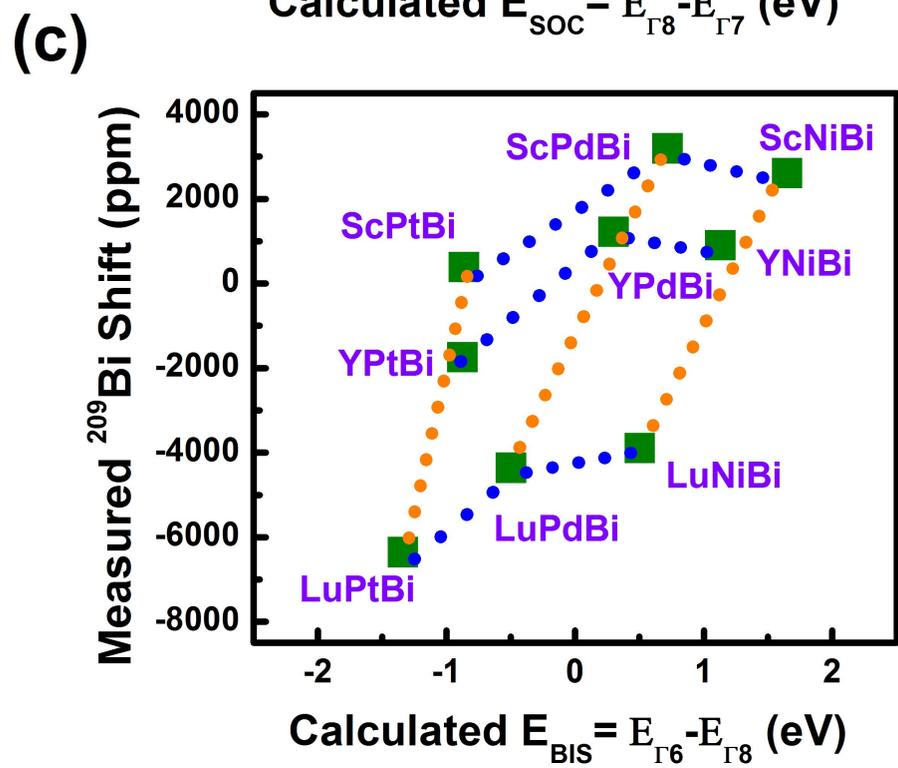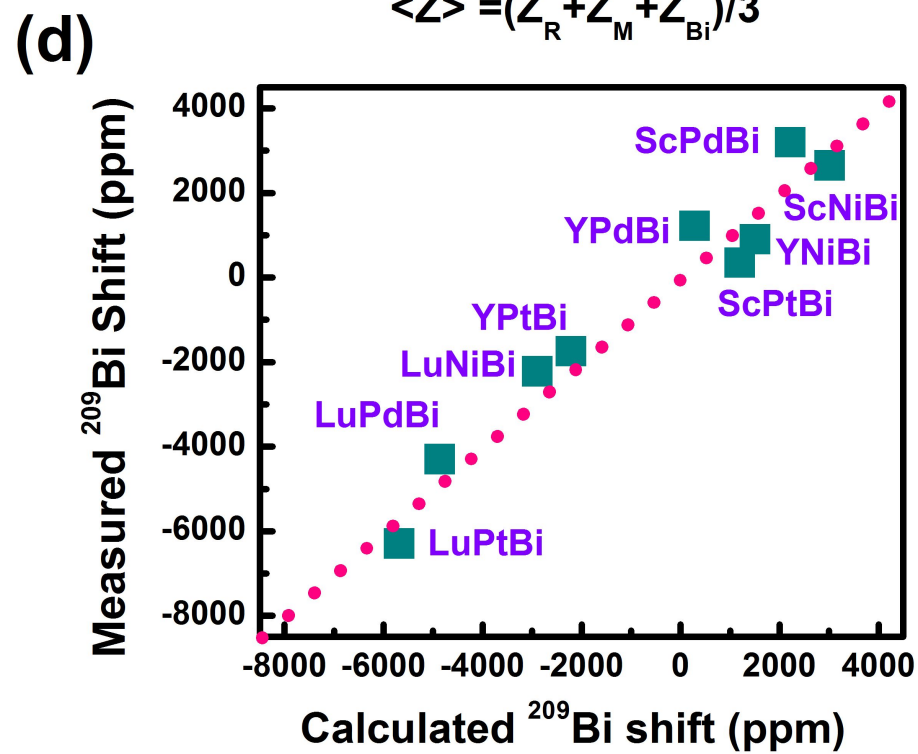

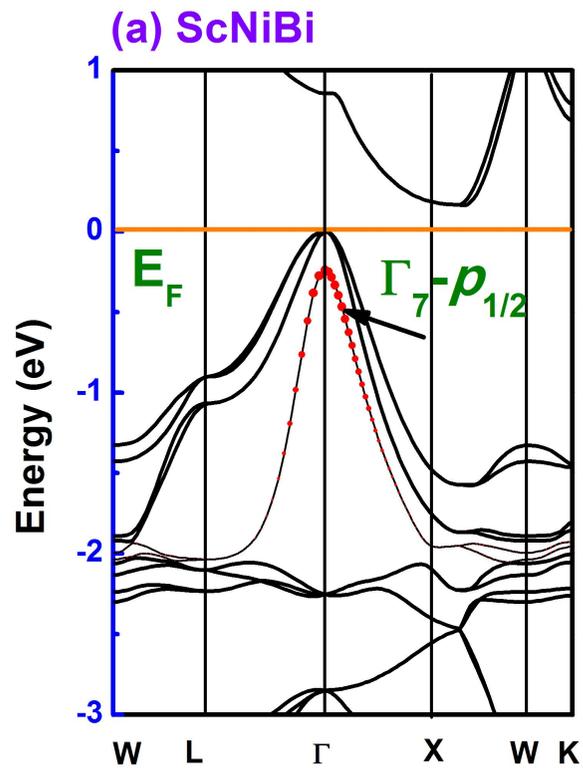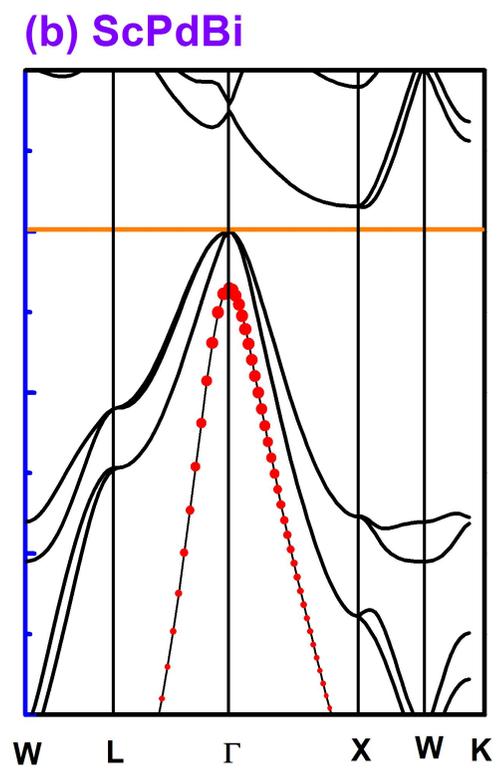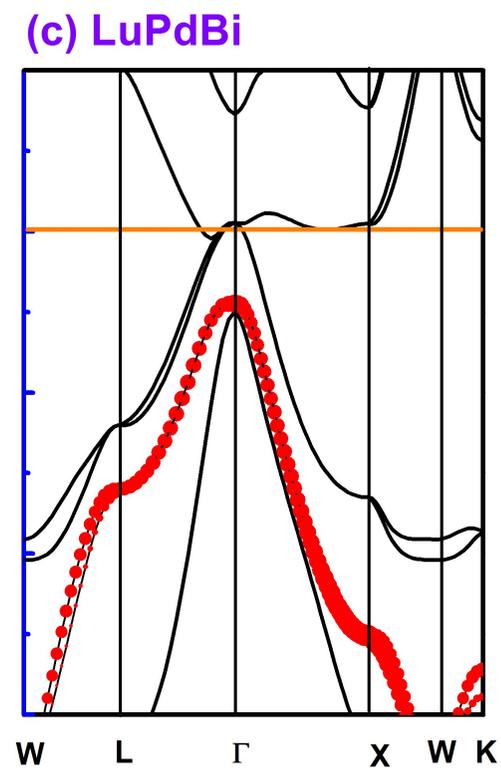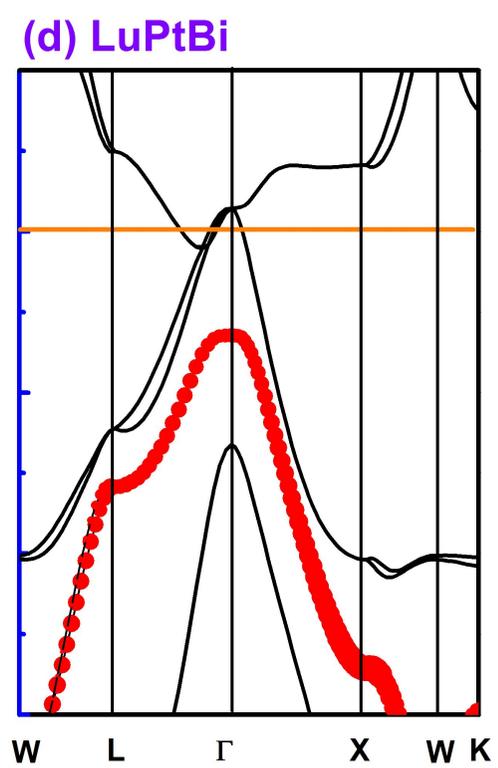